\documentclass[pra,twocolumn,amsmath,amssymb]{revtex4-1} %

\usepackage{epsfig,amsmath}
\usepackage{subfigure}
\usepackage{graphicx}
\usepackage{dcolumn}
\usepackage{stmaryrd}
\usepackage{mathrsfs}
\usepackage{pifont}
\usepackage{amsthm}
\usepackage{amssymb}
\usepackage{bm}
\usepackage{latexsym}
\usepackage{hyperref}
\usepackage{color}

\newcommand{\la}{\langle}
\newcommand{\ra}{\rangle}

\newcommand{\ti}{\tilde}
\newcommand{\ga}{\gamma}
\newcommand{\Ga}{\Gamma}

\newcommand{\da}{\dagger}
\newcommand{\De}{\Delta}

\newcommand{\si}{\sigma}

\newcommand{\om}{\omega}

\newcommand{\de}{\delta}

\newcommand{\non}{\nonumber}
\newcommand{\pa}{\partial}
\newcommand{\ket}[1]{\left\vert#1\right\rangle}

\def\jpb#1{{ J.\ Phys.\ B} {\bf#1}}

\def\pra#1{{ Phys.\ Rev. A\/} {\bf#1}}
\def\prb#1{{ Phys.\ Rev. B\/} {\bf#1}}

\def\prl#1{{ Phys.\ Rev.\ Lett.} {\bf#1}}
\def\pr#1{{ Phys.\ Rev.} {\bf#1}}

\def\pla#1{{ Phys.\ Lett. A\/} {\bf#1}}

\def\hyph{-\penalty0\hskip0pt\relax}

\begin{document}

\title{Random Control over Quantum Open Systems}

\author{Jun Jing$^{1,2}$, \footnote{[Email address]: jingjun@shu.edu.cn} C. Allen Bishop$^{3}$, and Lian-Ao Wu$^{2,4}$ \footnote{Correspondence to [Email address]: lianao\_wu@ehu.es}}

\affiliation{$^{1}$Institute of Theoretical Physics and Department of Physics, Shanghai University, Shanghai 200444, China\\ $^{2}$Department of Theoretical Physics and History of Science, The Basque Country University (EHU/UPV), PO Box 644, 48080 Bilbao Spain\\ $^3$GridCOM Technologies, Inc., San Diego, CA 92081, USA \\ $^4$Ikerbasque, Basque Foundation for Science, 48011 Bilbao Spain}

\date{\today}

\begin{abstract}
Parametric fluctuations or stochastic signals are introduced into the control pulse sequence to investigate the feasibility of random control over quantum open systems. In a large parameter error region, the out-of-order control pulses work as well as the regular pulses for dynamical decoupling and dissipation suppression. Calculations and analysis are based on a non-perturbative control approach allowed by an exact quantum-state-diffusion equation. When the average frequency and duration of the pulse sequence take proper values, the random control sequence is robust, fault-tolerant, and insensitive to pulse strength deviations and interpulse temporal separation in the quasi-periodic sequence. This relaxes the operational requirements placed on quantum control experiments to a great deal.
\end{abstract}


\maketitle

\section{Introduction}

Quantum technology allows us to organize and control the components of a complex system
governed by the laws of quantum physics \cite{Milburn}. Control lies at the heart of virtually every aspect of quantum science, and it is very difficult to achieve. A retrospective and representative study of control follows the spin echo effect; inhomogeneous broadening of a spin can be removed by applying a single $\pi$ inversion pulse halfway through the evolution 
\cite{Hahn}. Quantum control theory has matured over the last two decades, from single spin nuclear magnetic resonance experiments to large-scale manipulation during a quantum computation \cite{Nielsen}. 
As long as quantum technology has room to grow, quantum control will remain an active area of research.

Investigations based on the control of closed quantum system \cite{Rice,Shapiro1} have been extended into the open system regime \cite{Wu1} where active control sequences have been established to defeat the environmental dephasing and dissipation effects \cite{Leggett,Preskill,Lidar1,Breuer,Gardiner}. Mathematical control theorems have established the Lie-algebraic conditions required for strong analytic controllability \cite{HTC,Ramak} and the quantum limitation on control due to the dimensionality of the controlled subspace \cite{Shapiro2}. Quantum feedback control \cite{Wiseman} has been shown to stabilize \cite{Hope} or improve entanglement \cite{Yi} and rapidly prepare quantum states \cite{Combes} in the presence of control imperfections. Adiabatic evolution has been applied to quantum gate teleportation \cite{Bacon}, decoherence-free ground state dynamics \cite{Pekola}, and resonant passage with perfect fidelity \cite{Hu}. Exact state transmission can be achieved using a complete set of orthogonal states \cite{Wu2,Wu3} or nonuniform couplings \cite{Landahl} in a quantum spin network. In light of the quantum Zeno effect, decoherence of an open quantum system could be universally slowed down with ultra-fast modulation \cite{Kofman,Kurizki,Lidar2} and concatenated pulses \cite{Lidar4}. Many of these previous works have been assisted with dynamical decoupling (DD) by subjecting the target system to a series of open-loop, high-frequency, control transformations \cite{Viola,Uhrig,Liu,Lidar3,Gong,Wu4,Zhang}.

The control efficiency of DD is determined by the ratio of the characteristic timescale of the environmental correlation function and the period or quasi-period of the pulses. In the Markov environment, we are faced with a vanishing correlation time; control is impossible and any information contained within a quantum state is irreversibly lost. Whereas in a non-Markovian environmental, it is possible to drive a quantum state against the environmental influence with a properly configured control pulse sequence. Standard treatments invoke a delta-function idealization describing a zero-width pulse train of unbounded strength. Realistic experiments are not ideal; pulse strength is bounded, duration is finite, parametric fluctuations are inevitable. We can never completely eliminate stochastic quantum fluctuation and noise 
from the laboratory. In other words, the control mechanisms used in quantum control are generally not under total control themselves. In light of this reality, we are prompted to answer the following question: To what extent can a random control sequence be effective?

The randomness considered here is entirely different from the random decoupling schemes initiated by Viola et al. (see Ref. \cite{Viola1,Kern,Viola2}). The random DD considered there allows the control propagator to follow a random but known path on the group of rotating operations. Both the past control operations and the times at which they are applied are known, but the future control path is random. Our random control procedure assumes the occurrence of {\it unknown} stochastic fluctuations over pulse sequence parameters, including duration time, period, and strength. This investigation also differs from the control with no control procedure proposed by some of the authors of this work, see Ref. \cite{JW}. There, an out of control environment consisting of disordered external white noise was shown to suppress decoherence in a more ordered system. Here, we demonstrate the effectiveness of quantum state stabilization when irregular control pulses are used.

The rest of this paper is organized as follows. In Sec. \ref{mod}, we will introduce parameters with qualitative fluctuations and represent the random control model in the non-perburbative control approach based on the quantum-state-diffusion (QSD) equation. In Sec. \ref{dis}, fault-tolerant control will be demonstrated with fidelity calculations of a qubit system interacting with a dissipative environment. The results will be analyzed and discussed before concluding in Sec. \ref{con}.

\section{Random Control Model} \label{mod}

\begin{figure}[htbp]
\centering
\includegraphics[width=3.2in]{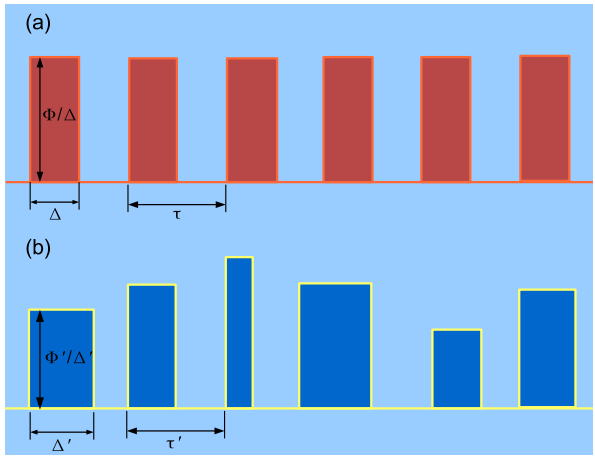}
\caption{The diagrams of (a) perfect pulse sequence with constant distance, duration, and strength (b) irregular pulse sequence with random distance, duration, and strength}\label{dia}
\end{figure}

A common approximation used in quantum control consists of an idealized sequence of delta functions of unbounded strength. This is actually the zero-order approximation to a non-perburbative control pulse sequence \cite{Jing,Wang}. We will instead use the more realistic pulse description consisting of a rectangular pulse sequence of finite width and strength. An illustration of a regularly spaced rectangular wave train is shown in Fig. \ref{dia}(a). This configuration allows us to parameterize experimental inaccuracies associated with the period $\tau$, temporal duration $\Delta$, and strength $\Phi/\Delta$. A typical example of an imperfect or out-of-order rectangular pulse sequence is shown in Fig. \ref{dia}(b). Notice the time-dependent strength, temporal duration, and quasi-period in this case. We will use these parameters to measure the fault-tolerance of a system to pulse sequence inaccuracies. The regularly spaced sequence appearing in Fig. \ref{dia}(a) should protect the quantum system from decoherence and relaxation. The robustness of our scheme can be portrayed by its fault-tolerance to fluctuations of three parameters. The quasi-period is defined as the time interval between the starting point of any pulse and that of its following pulse. The quasi-period will generally possess a random distribution. We can express these three fluctuating parameters as
\begin{equation}
X'=X+D_X{\rm Rand(-1,1)},
\end{equation}
where $X=\tau,\Delta,\Phi$, respectively, $D_X$'s are their individual deviation scales and ${\rm Rand(-1,1)}$ denotes a random number uniformly distributed between $-1$ and $1$. 
This is a realistic noise model $\hyph$ stochastic fluctuations naturally occur within the quantum measurement devices and during the pulse creation process. In the limit of a large sample number, $M[X']=X$, where $M[\cdot]$ denotes the statistical ensemble mean. Each random series is instantaneously generated and mutually independent. Furthermore, the random values are not shared between $\tau,\Delta$, and $\Phi$ simultaneously. More importantly, they are unknown during our control scheme. Without loss of generality, we require $\Delta+D_\Delta<\tau-D_\tau$ for each rectangular pulse to avoid unnecessary confusion.

We employ an exact stochastic Schr\"odinger equation, the QSD equation, to investigate this stochastic control process. This allows us to include the dynamical decoupling control function directly into the microscopic quantum model. For an arbitrary model (setting $\hbar=1$):
\begin{equation}
H_{\rm tot}=H_{\rm sys}(t)+\sum_k(g_k^*L{\hat{a}}_k^\da+h.c.)+\sum_k\om_k{\hat{a}}_k^\dag {\hat{a}}_k
\end{equation}
where ${\hat{a}}_k^\dag$ and ${\hat{a}}_k$ respectively denote the creation and annihilation operators for the $k$-th mode in the bosonic bath. In the interaction picture with respect to $\sum_k\om_k{\hat{a}}_k^\dag {\hat{a}}_k$, an exact QSD equation generally reads:
\begin{eqnarray}\non
\pa_t\psi_t(z^*)&=&-i[H_{\rm sys}(t)+iLz_t^*-iL^\dag\bar{O}(t,z^*)]\psi_t(z^*) \\ \label{qsd}
&=&-iH_{\rm eff}(t)\psi_t(z^*).
\end{eqnarray}
Here $H_{\rm sys}(t)$ is the system Hamiltonian that could absorb an arbitrary control pulse function. 
We have dropped the ket notation for $\psi_t(z^*) \equiv \ket{\psi_t(z^*)}$. $L$ is the coupling operator of the system with the environment. The stochastic system wave function is constructed by $\psi_t(z^*)\equiv\la z_1|\la z_2|\cdots\la z_k|\cdots|\Psi(t)\ra$, where $|\Psi(t)\ra$ is the total wave-function for the system plus environment and $|z_k\ra$ is the random Bargmann coherent state for mode-$k$. $z_t^*\equiv-i\sum_kg_k^*z_k^*e^{i\om_kt}$ is the environmental Gaussion noise process satisfying $M[z_t^*]=M[z_t^*z_s^*]=0$ and $M[z_tz_s^*]=\alpha(t,s)$ and $\alpha(t,s)$ is the environmental correlation function. $\bar{O}(t,z^*)$ includes system operators and along with the environmental noises which satisfy the consistency conditions \cite{Diosi1}. These conditions are used to describe the environmental influence without invoking the master equation: $\bar{O}(t,z^*)\psi_t(z^*)=\int_0^tds\alpha(t,s)O(t,s,z^*)\psi_t(z^*)$ with  $O(t,s,z^*)\psi_t(z^*)\equiv\frac{\de\psi_t(z^*)}{\de z_s^*}$. The states $\psi_t(z^*)$ of the system correspond to a particular "configuration" $z^*$, thus the system density matrix is recovered by $\rho_t=M[|\psi_t(z^*)\ra\la\psi_t(z^*)|]$.

We note that identical results would hold for our dynamical-decoupling control process if we had used a density matrix description instead. The QSD equation is exact, it is based solely on the microscopic model without any approximation. Therefore our random control approach is a kind of non-perturbative control theory. Another advantage of employing the QSD equation is that it allows one to take the memory effect of the non-Markovian environment into account. Many recent experiments are non-Markovian and could be represented by the correlation function $\alpha(t,s)$. Instead of the Markov Wiener process, which satisfies $M[z_tz_s^*]=\Ga\delta(t,s)$, where $\Ga$ is the system-environment coupling strength, we consider here the Ornstein-Uhlenbek process of environmental noise: $M[z_tz_s^*]=\frac{\Ga\ga}{2}e^{-\ga|t-s|}$, where $1/\ga$ is proportional to the memory time of the environment. When $\ga\rightarrow\infty$, this non-Markovian process is reduced to the Markov limit asymptotically.

\section{Fault-tolerant Control and Discussion} \label{dis}

Consider the fidelity control over one two-level system (qubit) which is under the dissipative influence from the environment. Then the original system Hamiltonian and coupling operator read $H_{\rm sys}^{(0)}=\om\si_z/2$ and $L=\si_-$. The pulse sequence, regular or random, contributes to the system energy shift $\om\rightarrow E(t)=\om+c(t)$, where $c(t)=\Phi'/\Delta'$ during the "on" time which lasts $\Delta'$ and subsequently $c(t)=0$ during the "off" time of duration $\tau'-\Delta'$ during an interval of instantaneous quasi-period $\tau'$.

\begin{figure}[htbp]
\centering
\includegraphics[width=3.2in]{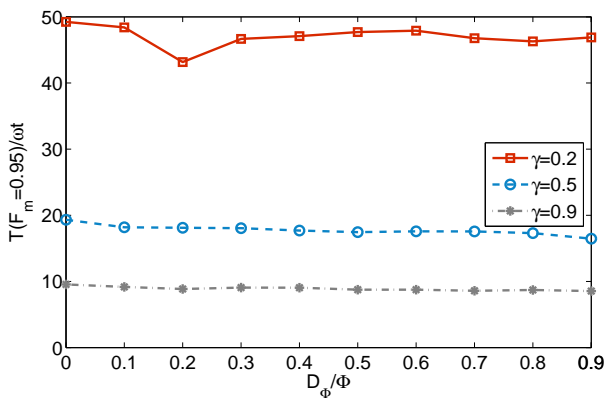}
\caption{The moment $T$ when the average fidelity of a single qubit reduces to $0.95$ from unity as a function of the pulse strength fluctuation scale under different $\gamma$. We choose $\Phi=0.2\omega$, $\tau=0.02\om t$, $\Delta=0.4\tau$ and $\Ga=\om$.}\label{phi}
\end{figure}

\begin{figure}[htbp]
\centering
\includegraphics[width=3.2in]{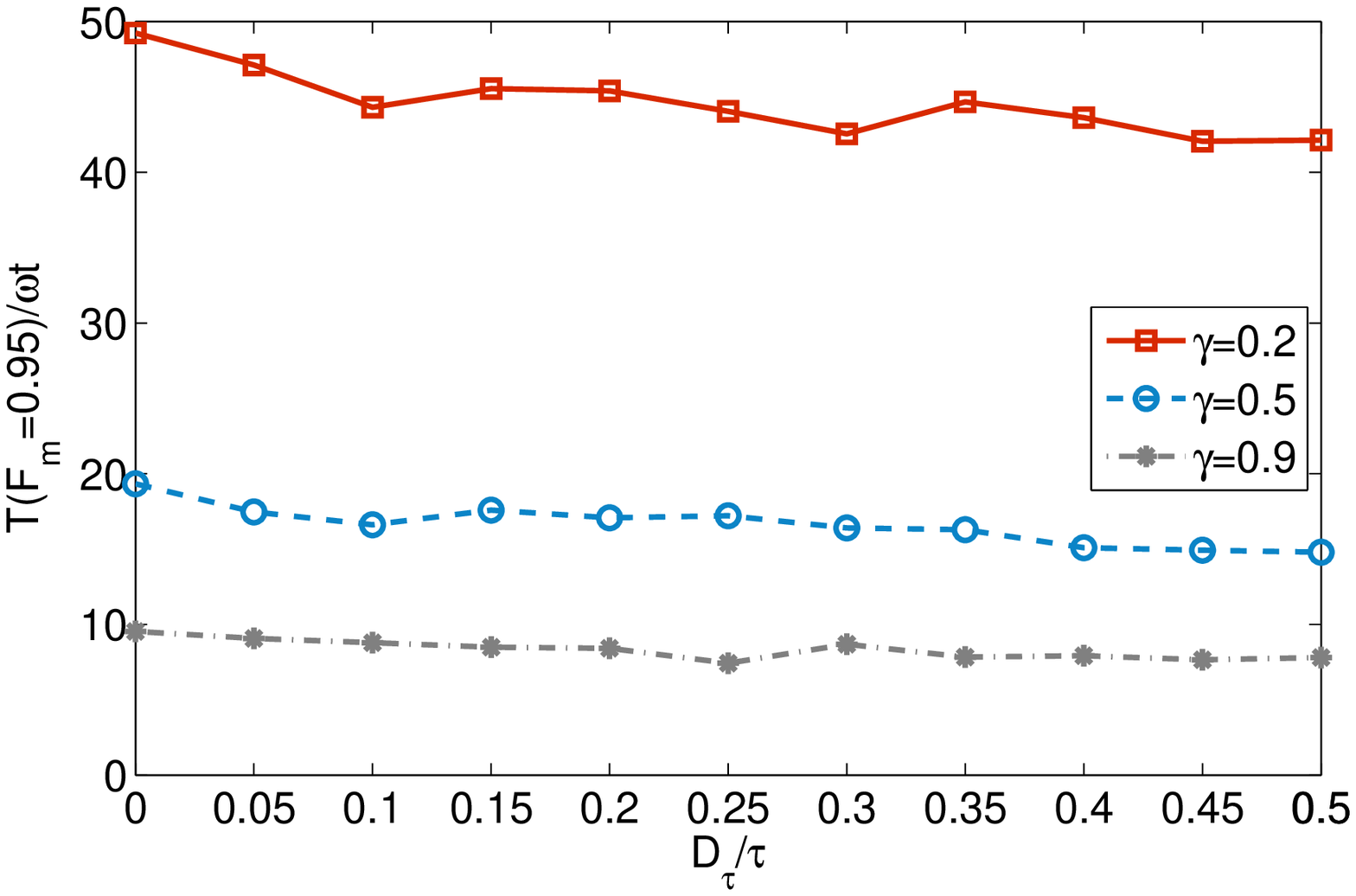}
\caption{The moment $T$ when the average fidelity of a single qubit reduces to $0.95$ from unity as a function of the period fluctuation scale under different $\gamma$. We choose $\Phi=0.2\omega$, $\tau=0.02\om t$,  $\Delta=0.4\tau$ and $\Ga=\om$.}\label{tau}
\end{figure}

\begin{figure}[htbp]
\centering
\includegraphics[width=3.2in]{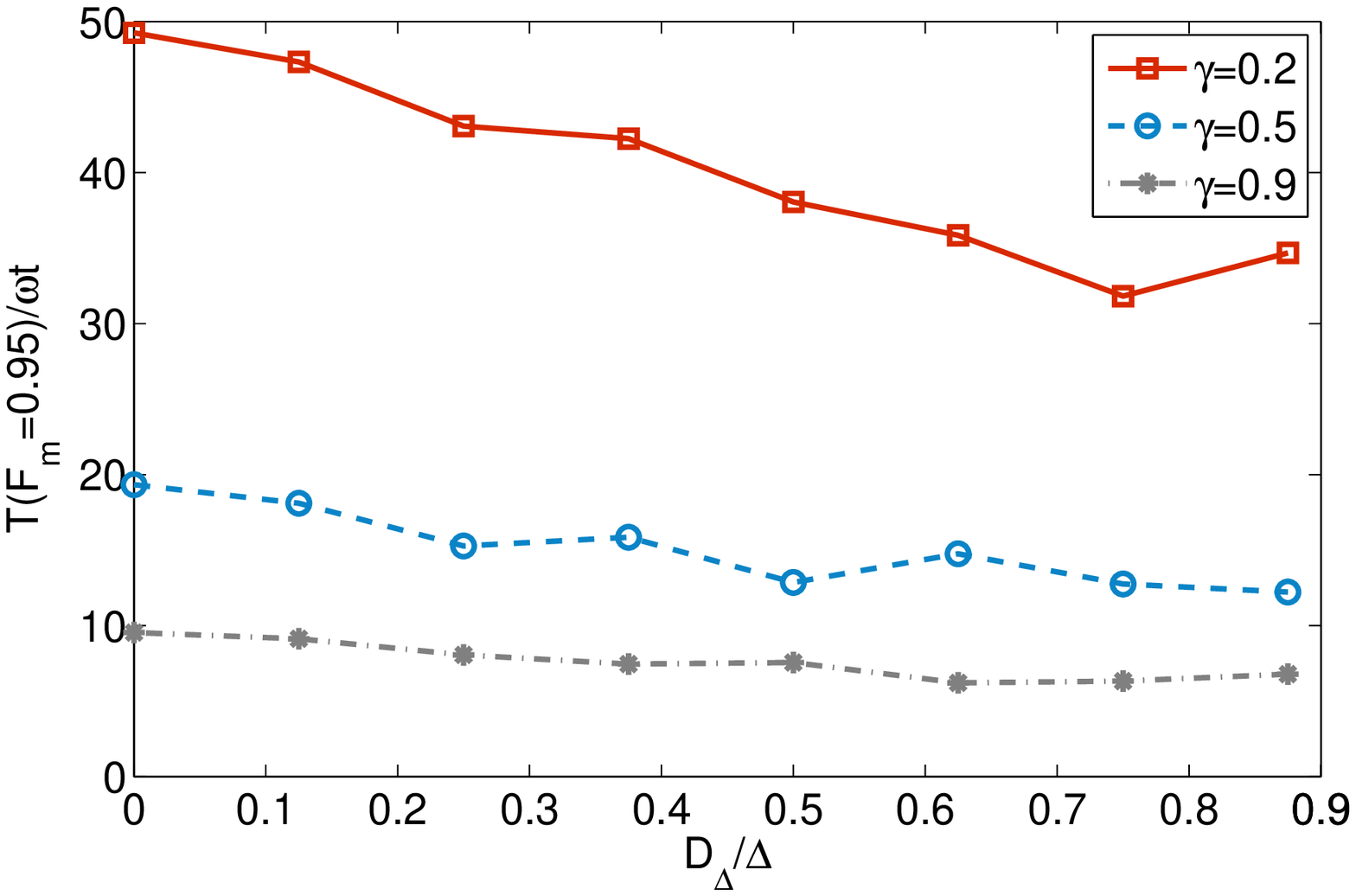}
\caption{The moment $T$ when the average fidelity of a single qubit reduces to $0.95$ from unity as a function of the temporal duration fluctuation scale under different $\gamma$. We choose $\Phi=0.2\omega$, $\tau=0.02\om t$, $\Delta=0.4\tau$ and $\Ga=\om$.}\label{delta}
\end{figure}

For this model, the exact O-operator is found to be $Q(t)\si_-$ from the equation of motion for the O-operator:
\begin{equation}
\pa_tO(t,s,z^*)=[-iH_{\rm eff}(t), O(t,s,z^*)]-L^\da\frac{\de O(t,z^*)}{\de z_s^*},
\end{equation}
which can be obtained using the fact $\frac{\pa}{\pa_t}\frac{\de\psi_t(z^*)}{\de z_s^*}=\frac{\de}{\de z_s^*}\frac{\pa\psi_t(z^*)}{\pa_t}$. Then the effective Hamiltonian in Eq.~(\ref{qsd}) reads: \cite{Diosi1,Diosi2}
\begin{equation}
H_{\rm eff}(t)=\frac{\omega+c(t)}{2}\si_z+iz_t^*\si_--iQ(t)\si_+\si_-,
\end{equation}
where $Q(t)$ satisfies an ordinary differential equation
\begin{equation}\label{dQt}
\pa_tQ(t)=\frac{\Ga\ga}{2}+[-\ga+i\om+ic(t)]Q(t)+Q^2(t),
\end{equation}
and the boundary condition is $Q(0)=0$. The system is initialized to $|\psi_0\ra=\mu|1\ra+\nu|0\ra$, $|\mu|^2+|\nu|^2=1$. The fidelity is defined to be $\mathcal{F}(t)\equiv\la\psi_0|\rho_t|\psi_0\ra$.

The initial state evolves under dissipation and control. At time $t$ the initialization survives with probability:
\begin{eqnarray}\non
\mathcal{F}(t)&=&1-|\mu|^2-(|\mu|^2
-2|\mu|^4)e^{-2\int_0^tds\mathcal{R}[Q(s)]}\\ \label{qubit0} &+&2|\mu|^2(1-|\mu|^2)\mathcal{R}[e^{-\int_0^tdsQ(s)}],
\end{eqnarray}
where $\mathcal{R}[\cdot]$ takes the real part of the following function. We average the fidelity over all possible pure state configurations and find
\begin{equation}\label{qubit}
\mathcal{F}_m(t)=\frac{1}{2}+\frac{e^{-2\int_0^tds\mathcal{R}[Q(s)]}}{6} +\frac{\mathcal{R}[e^{-\int_0^tdsQ(s)}]}{3}.
\end{equation}

We will define the efficiency of regular and random control in terms of the time interval associated with a fidelity drop from unity to $0.95$. The fidelity is functionally dependent on the bath memory coefficient $\ga$. We illustrate the relationship between the memory function and the randomized control parameters $\Phi$, $\tau$, and $\De$ separately in Figs. \ref{phi}, \ref{tau}, and \ref{delta}. Each calculation determined the randomization effects inherent to a specific control parameter while the remaining parameters remained fixed in a dummy state. The efficiency was then evaluated with respect to the ensemble average of these dummy states.

The horizontal axes appearing in Figs. \ref{phi}, \ref{tau}, and \ref{delta} represent the ratio of the stochastic fluctuation scale $D_X$ to the mean value of the corresponding parameter $X$; $X=\Phi$, $\tau$, and $\De$, respectively. The fluctuations vanish at the origin of each of these coordinate systems. This is the condition for regular control, hence, for a fixed $\ga$, the relative efficiency of random and regular control appears in reference to the intersection of the curve and the vertical axis. These results should also be compared to those which naturally emerge in the absence of control: $T(\mathcal{F}_m=0.95)=1.42\om t$, $0.87\om t$, and $0.65\om t$ when $\ga=0.2$, $05$ and $0.9$, respectively. Thus it is evident that both random and regular control greatly enhance the survival time of the target state as it evolves under open system dynamics; order-of-magnitude improvements are found even in the worst situations considered. 

The efficiency has a varying sensitivity to deviations in the control parameters. This behavior is illustrated in Figs. \ref{phi}, \ref{tau}, and \ref{delta}. For example, deviations 
in the pulse strength hardly effect the value of $T(\mathcal{F}_m=0.95)$. Deviations in the period of the pulse sequence have a small impact on the efficiency as well, though a modest decline in $T(\mathcal{F}_m=0.95)$ tends to occur with increasing $D_\tau/\tau$. However, the performance of the control sequence becomes sensitive to deviations in $\Delta$ as the bath approaches the strong non-Markovian regime. This sensitivity is exemplified in Fig. \ref{delta} by the $40\%$ reduction of $T$ over the evaluation scale $5/4\De<\tau'<15/4$ with $\ga=0.2$, a typical strong non-Markovian condition. These results indicate the effectiveness of noisy quantum control. The relative performance of regular and random control depends on the environmental coefficient $\ga$, improving as the environment becomes more Markovian. For near-Markovian processes, little distinction can be made between completely regular and completely random control.

The common thread among these three figures is the memory effect of the environment. In terms of the fidelity expression in Eq.~(\ref{qubit}), strong control requires the exponential function $\bar{Q}(t)\equiv\exp[-\int_0^tdsQ(s)]$ to approach unity in the presence of $c(t)$. In absence of control, 
\begin{eqnarray}\non
\bar{Q}(t)&=&e^{-\ti{\ga}t}\sqrt{1-\frac{\ti{\ga}^2}{2\Ga\ga}}\cos\bigg[\frac{\sqrt{2\Ga\ga-\ti{\ga}^2}}{2}t
\\ \label{barQ} &-&\arctan(\frac{\ti{\ga}}{\sqrt{2\Ga\ga-\ti{\ga}^2}})\bigg]
\end{eqnarray}
where $\ti{\ga}=\ga-i\om$. A straightforward derivation shows that when $\ga$ is dominant in Eqs.~(\ref{dQt}) and (\ref{barQ}), the absolute vale of $\bar{Q}(t)$  approximately decays in an exponential way with time. Thus the control pulse is not able to dynamically decouple the open system for dominantly large $\ga$, i.e. within the near-Markovian environment. On the contrary, small $\ga$ implies a long memory and a slower characteristic decay of the quantum state. Coherence can be preserved in this case with a control pulse of proper frequency. Appropriate pulse sequences could roughly wash out the dissipation influence.

In summary, we have found that standard quantum control can be quite effective when fluctuations occur in the pulse sequence. The performance is largely insensitive to fluctuations in pulse strength and sequence period. It is, however, quite sensitive to several other factors. The greatest influence, to no surprise,  comes from the structure of the environmental noise process itself. A non-Markovian environment is required for effective control in both cases, regular and random. Secondly, the control sequence requires a sufficiently fast pulse series, i.e. a sufficiently small average value of $\tau$. Dissipation effects from the system-environment interaction create disordered system dynamics, these effects can be neutralized if a sufficiently large number of control photons interact with the system during the characteristic correlation time of the environment. Hence, the required pulse rate depends on the fixed bath correlation time, a time that is inversely proportional to $\ga$. 
 
\begin{figure}[t]
\centering
\includegraphics[width=3.2in]{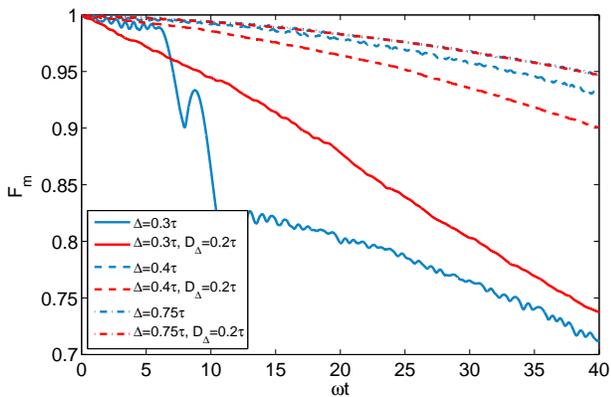}
\caption{The time evolution of the average fidelity of a single qubit under regular control (blue lines) and under random control (red lines) with different $\Delta$. We choose $\ga=0.3$, $\Phi=0.2\omega$, $\tau=0.02\om t$ and $\Ga=\om$.}\label{dyndelta}
\end{figure}

\begin{figure}[t]
\centering
\includegraphics[width=3.2in]{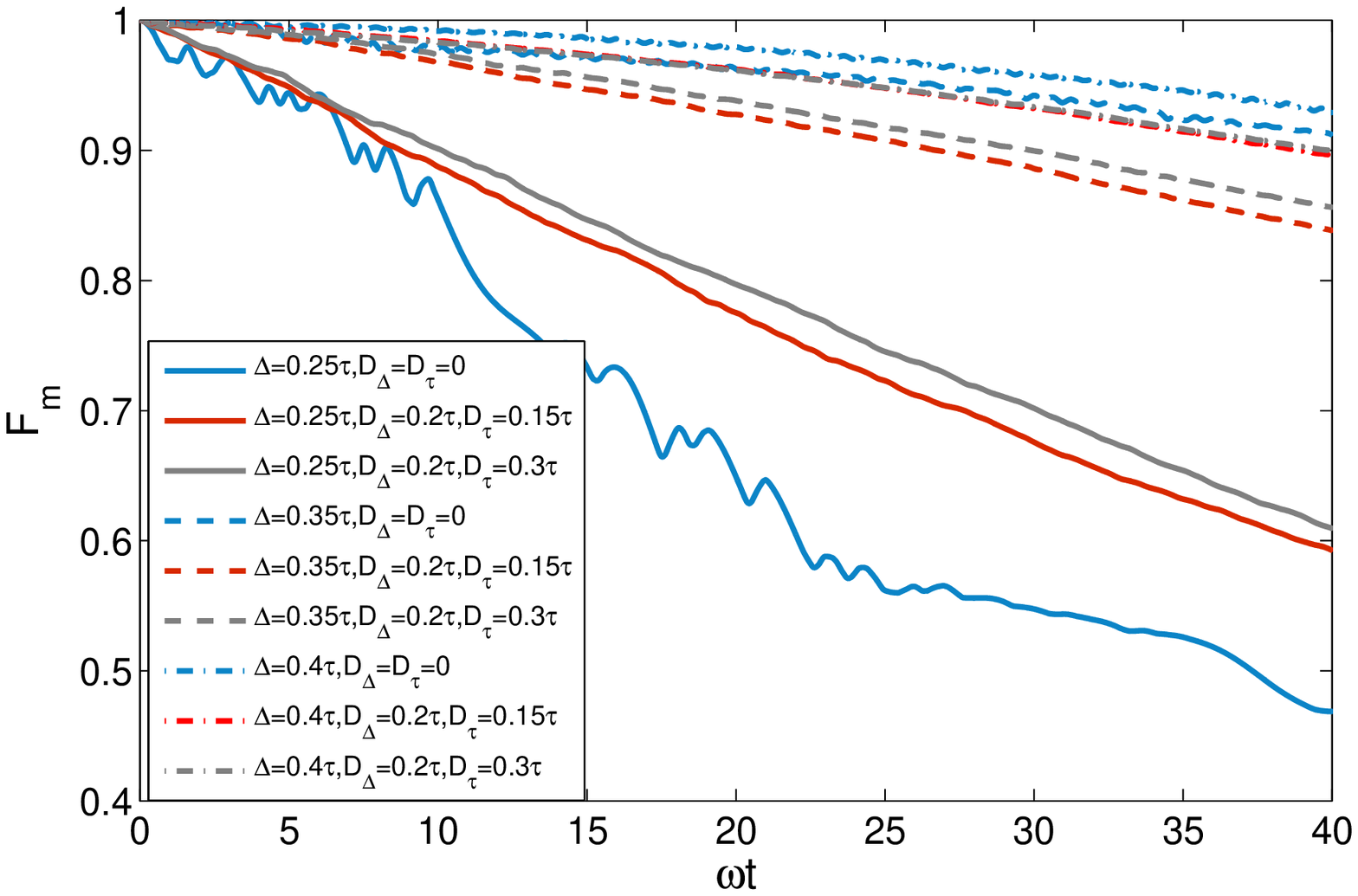}
\caption{The time evolution of the average fidelity of a single qubit under regular control (blue lines) and under random control (red and gray lines) with different $\Delta$, $D_\Delta$ and $D_\tau$. We choose $\ga=0.3$, $\Phi=0.2\omega$, $\tau=0.02\om t$ and $\Ga=\om$.}\label{dyndeltatau}
\end{figure}

A third condition, which has not been discussed thus far, concerns the ratio between the pulse width and pulse periodicity. This dependence is clearly illustrated in Fig. \ref{dyndelta} where comparisons in the fidelity evolution are made between regular and stochastic control for various $\De/\tau$. Perhaps the most striking comparison results when $\De/\tau=0.3$ $\hyph$ random control can actually outperform regular control in some occassions. To be fair, we concede that regular control is more favorable towards the beginning of this example evolution. As the interpulse separation deminishes, e.g. $\De=0.75\tau$, it becomes difficult to discriminate the curves and the fidelity remains roughly the same in either case. Moreover, in Fig. \ref{dyndeltatau}, similar patterns emerge when we examine inter-parameter dependencies between $\De$ and $\tau$. Both of these figures correspond to the same average $\De$ and $\tau$. Again, we find examples where random control permits longer survival times than the regular case for long evolutionary periods. However, in this case random control outperforms regular control in early evolutionary intervals as well. Over the entire evaluation interval, random control yields a fidelity greater than $0.9$ when the ratio between the average values of $\De$ and $\tau$ is optimized ($\De/\tau\geq0.4$). Although the optimized random controls yield slightly lower fidelities than the regular controls for these average values, they could adequately approximate the ideal regular sequence if the performance requirements are not prohibitively restrictive.
\begin{figure}[htbp]
\centering
\includegraphics[width=3.2in]{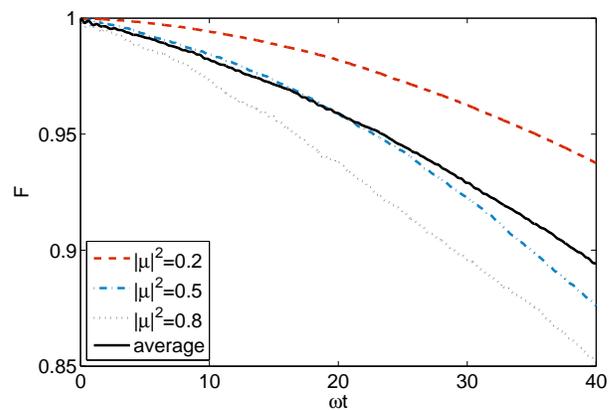}
\caption{The time evolution of the fidelity of a single qubit under random control with different initial states indicated by population $|\mu|^2$. We choose $\ga=0.3$, $\Phi=0.2\omega$, $\tau=0.02\om t$,  $\Delta=0.4\tau$, $D_\Delta=0.2\tau$, $D_\tau=0.2\tau$ and $\Ga=\om$.}\label{dynmu}
\end{figure}

The effectiveness of the random control sequence is fairly state-independent. In Fig. \ref{dynmu}, we illustrate the similarity in performance for several initial qubit configurations. These configurations are defined in terms of the initial population 
$|\mu|^2$ of the upper level of the open qubit system. We plot the fidelity as a function of $|\mu|^2$ rather than $\mu$ in light of Eq.~(\ref{qubit0}). Curves associated with a particular $|\mu|^2$ represent the common result of a group of quantum states with the same upper level population. We find a slight decrease in control efficiency with increasing $|\mu|^2$. Nevertheless, random control provides robust and reliable stabilization against damping effects for all possible initializations. In the short time regime, the efficiency of random control is nearly the same for every possible initial state.

\section{Conclusion}\label{con}

We have introduced a general unknown stochastic disturbance into the idealized quantum control model to account for imperfect laboratory control. An idealized control sequence consists of a sequence of identical, equally spaced pulses. With high precision, we can control the average values of the parameters which characterize the sequence, but we cannot control the errors which occur on them or predict any occurrence in advance. Using the quantum-state-diffusion equation, we have analyzed the performance of a realistic quantum control sequence prepared with random laboratory parametric fluctuations and find it to be effective in controlling the dynamics of an open quantum system over a wide range of non-Markovian processes.

The control efficiency, which is measured by the average fidelity over initial pure states, is found to be dependent on the average frequency of the control pulse, the environmental memory, and the ratio of the average pulse duration time $\De$ to the  period $\tau$. When the mean values of $\De$ and $\tau$ are optimized, random control sequences can adequately approximate the ideal regular sequence if the performance requirements are not prohibitively restrictive. This will alleviate the experimental requirements placed on pulse sequence generation in quantum control to a great extent.

\acknowledgments
We acknowledge grant support from the NSFC No. 11175110, an Ikerbasque Foundation Startup, the Basque Government (grant IT472-10), and the Spanish MICINN (Project No. FIS2009-12773-C02-02).

\end{document}